\documentclass[10pt, conference, letterpaper]{IEEEtran}
\IEEEoverridecommandlockouts
\usepackage{cite}
\usepackage{amsmath,amssymb,amsfonts,amsthm}
\usepackage{algorithm}
\usepackage{algorithmic}
\usepackage{graphicx}
\usepackage{textcomp}
\usepackage{xcolor}
\usepackage{bm}
\def\BibTeX{{\rm B\kern-.05em{\sc i\kern-.025em b}\kern-.08em
    T\kern-.1667em\lower.7ex\hbox{E}\kern-.125emX}}

\newtheorem{property}{Property}

\DeclareMathOperator*{\argmin}{arg\,min}

\begin{document}

\setlength{\textfloatsep}{3pt}
\setlength{\dbltextfloatsep}{5pt}
\setlength{\floatsep}{2pt}
\setlength{\dblfloatsep}{2pt}

\setlength{\abovedisplayskip}{3pt}
\setlength{\belowdisplayskip}{3pt}
\setlength{\abovedisplayshortskip}{3pt}
\setlength{\belowdisplayshortskip}{3pt}

\title{Workflow-Aware Communication and Computation Scheduling for Mobile Edge AI Agents}

\author{
  \IEEEauthorblockN{
    Xingqiu He,
    Chaoqun You,
    Feng Wang,
    Zihan Chen,
    Shuai Wang,
    Jihong Park }

}

\maketitle

\begin{abstract}
AI agents are increasingly deployed in mobile edge environments to autonomously handle diverse tasks through multi-step workflows. Unlike conventional dependency-aware tasks with predefined dependency structures, agent workflows are dynamically generated based on intermediate execution results. To minimize the weighted sum of job completion latencies, this paper models partially revealed agent workflows using dynamic workflow graphs and formulates an online scheduling problem that jointly optimizes computation placement, wireless resource block allocation, and edge computing resource allocation. To address the uncertainty of future workflow evolution, we first transform the weighted completion time objective into an equivalent per-slot holding cost and characterize the optimal policy through a Bellman equation. We then construct a tractable look-ahead cost-to-go surrogate based on the revealed workflow and the estimated remaining latencies of active jobs. The proposed scheduling algorithm leverages the surrogate to determine computation placement and uses the resulting operation criticality to allocate wireless and edge computing resources online. 
Simulation results show that the proposed algorithm outperforms the benchmark schemes across diverse scenarios.

\end{abstract}


\section{Introduction}
Recent advances in large language models (LLMs) have enabled AI agents to reason, 
take actions, and invoke external tools~\cite{wang2024agentsurvey,yao2023react,schick2023toolformer,qin2024toolllm}. 
Such agents are increasingly deployed in interactive and mobile environments~\cite{liu2024agentbench,wang2024mobileagentv2,rawles2025androidworld} 
to support applications requiring both environmental awareness and intelligent decision-making. 
Unlike a single model query, an agent job is typically executed through a multi-step workflow 
consisting of operations such as sensing, preprocessing, data transmission, model inference, and tool invocation~\cite{shen2023hugginggpt}. 
These communication and computation operations often exhibit dependency constraints,
where an operation becomes ready only after all its predecessors are completed. 

Different from conventional dependency-aware tasks with known dependency structures, an agent workflow is dynamically generated based on intermediate execution results. 
Consequently, an online scheduler can only observe the revealed portion of the workflow and must allocate resources before future operations become available. 
Moreover, executing the entire workflow locally is often infeasible due to the limited computational capability of mobile devices. 
Although edge and cloud computing provide more powerful computational resources, offloading operations introduces additional transmission delays and resource contention. 
Therefore, the job completion latency is jointly determined by computation placement and the allocation of shared communication and computing resources~\cite{mao2017survey,li2020edgeai}.

To improve user experience, it is essential to optimize resource allocation to minimize the completion latency of agent jobs. 
However, designing an online scheduling algorithm for such workflows faces the following three challenges. 
First, the scheduler must evaluate the long-term impact of each decision based on an incomplete workflow, without knowing future operations and their dependencies. 
Second, communication and computation are tightly coupled through workflow dependencies: 
wireless transmission determines when downstream computation operations become ready, 
while computation placement determines both their processing rates and the additional data transmissions required along the workflow. 
Third, concurrent jobs compete for shared wireless and edge computing resources, whose availability varies over time due to channel fluctuations and background traffic. 
These factors make computation placement, wireless resource allocation, and edge computing allocation highly coupled.

In this paper, we investigate joint communication and computation scheduling for dynamically revealed mobile edge agent workflows. 
We model each partially revealed workflow as a dynamic workflow graph and formulate an online scheduling problem that 
minimizes the weighted sum of job completion latencies. 
To characterize how current decisions affect this objective, 
we first transform the weighted completion-latency objective into an equivalent per-slot holding cost and 
derive the corresponding Bellman equation. 
Since the exact cost-to-go cannot be evaluated due to unknown future workflow and resource states, we construct a tractable surrogate based on the revealed workflow and the estimated remaining latency of each active job. 
The proposed surrogate assigns smaller weights to holding costs predicted to occur farther in the future and 
incorporates a correction term to capture the state transitions caused by operation completion. 
Based on the resulting operation criticality, 
the proposed scheduler jointly determines computation placement and resource allocation for communication and computation operations in each slot.

The main contributions are summarized as follows.
\begin{itemize}
    \item We introduce a dynamic workflow graph to model agent workflows whose operations and dependencies are gradually revealed during execution.
    Based on this model, we formulate an online scheduling problem that jointly optimizes computation placement, wireless RB allocation, 
    and edge computing resource allocation to minimize the weighted sum of job completion latencies.
    \item We transform the weighted completion time objective into an equivalent holding-cost objective and construct a look-ahead cost-to-go surrogate based on the revealed workflow.
      By combining the workload gradient of the surrogate with a completion correction term, we develop an online scheduler that accounts for both downstream workflow dependencies and state transitions induced by operation completion.
    \item We develop an event-driven simulator with multiple workflow structures and time-varying communication and computing resources.
      Numerical results demonstrate that the proposed algorithm outperforms three benchmarks across diverse scenarios.
\end{itemize}

\section{Related Work}
\textbf{Dependency-Aware Task Scheduling.}
Scheduling tasks with precedence dependencies has been extensively studied in heterogeneous and edge computing.
HEFT assigns the operations of a known DAG to heterogeneous processors according to their upward ranks~\cite{topcuoglu2002heft}.
PEFT further uses an optimistic cost table to incorporate downstream execution costs into task ranking and processor selection~\cite{arabnejad2014peft}.
In edge computing networks, computation offloading further couples task execution with wireless and computing resource allocation~\cite{mao2017survey}.
Dependency-aware studies optimize offloading and scheduling across device, edge, and cloud resources~\cite{chen2022dependency,liu2023dependent}.
Recent online approaches further update task priorities as multiple applications arrive and resource conditions change~\cite{liu2026multidag}.
These methods account for resource heterogeneity and workflow dependencies, and some adapt their decisions online.
However, the complete set of operations and dependencies of each admitted application is available before scheduling begins.

\textbf{Resource Scheduling for Edge AI.}
Resource scheduling for edge AI has mainly focused on accelerating DNN inference under limited communication and computing resources.
Neurosurgeon~\cite{kang2017neurosurgeon}, JointDNN~\cite{eshratifar2021jointdnn}, and Edge AI~\cite{li2020edgeai} partition a DNN between mobile devices and infrastructure resources according to computation and communication conditions.
DADS further adapts DNN partitioning between edge and cloud resources to changing network conditions~\cite{hu2019dads}.
Beyond device--server partitioning, CoEdge distributes inference workloads across heterogeneous edge devices according to their computing capabilities and network conditions~\cite{zeng2021coedge}.
Jellyfish jointly selects data and DNN adaptations and batches inference requests under dynamic wireless latency~\cite{nigade2022jellyfish}, whereas Ekya allocates edge GPU resources between inference and model retraining~\cite{bhardwaj2022ekya}.
These studies provide effective resource scheduling mechanisms for AI inference, but their operation graphs are determined by fixed DNN architectures or predefined service pipelines. 
Autellix instead targets dynamically generated LLM-agent programs and introduces ATLAS, which prioritizes programs based on their attained service~\cite{luo2025autellix}. 
In contrast, agent workflows may interleave communication, heterogeneous computation, and tool invocation operations that are gradually revealed during execution.

\section{System Model and Problem Formulation} \label{sec:system_model}
In this section, we first introduce the dynamic agent workflow model and the wireless edge resource model,
then formulate the online workflow-aware scheduling problem.

\begin{figure}[t]
\centering
\includegraphics[width=\linewidth]{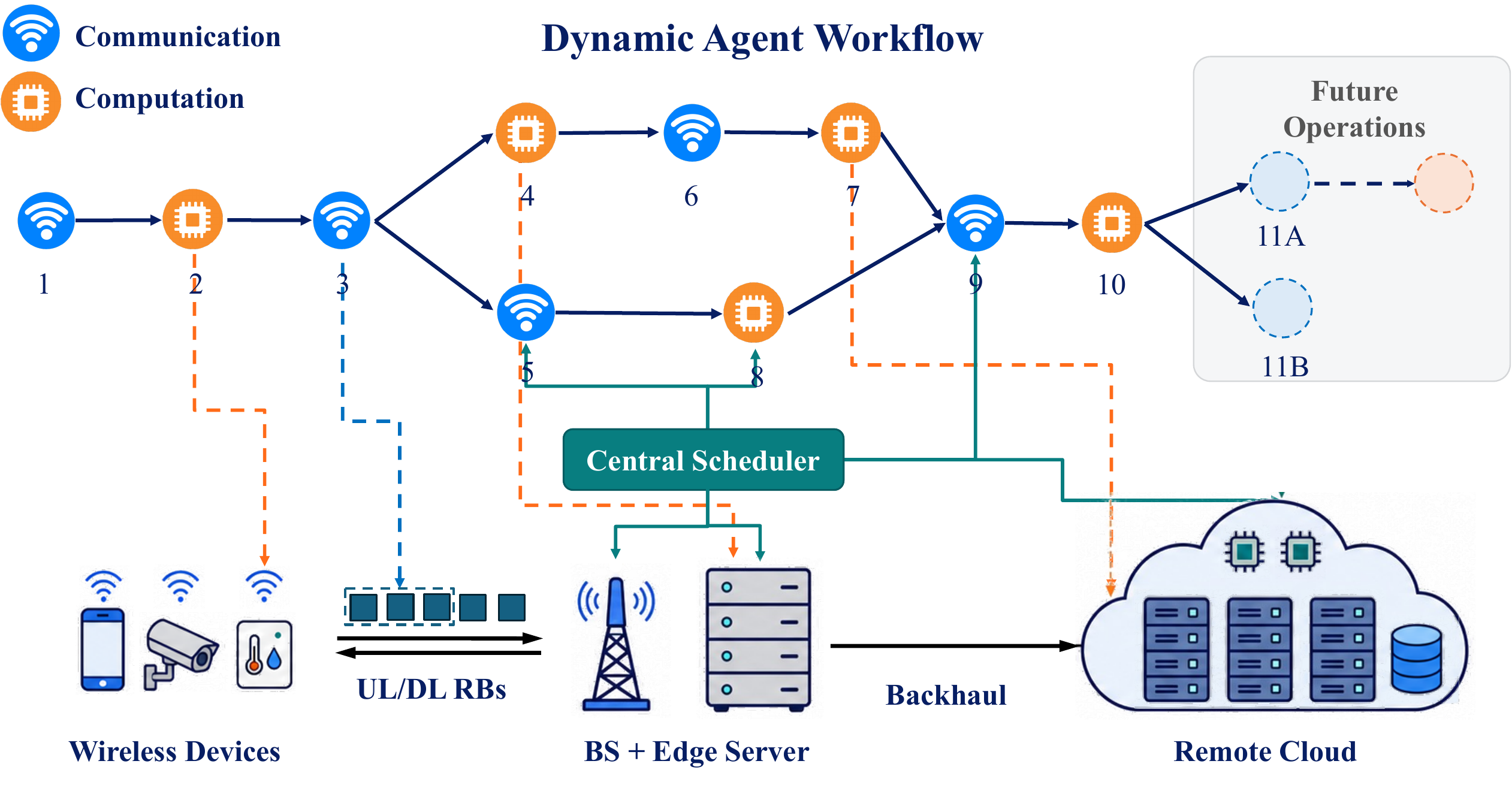}
\caption{Illustrative example of the considered system.}
\label{fig:system}
\end{figure}

\subsection{Dynamic Agent Workflow Model} \label{subsec:workflow_model}
As shown in Fig.~\ref{fig:system}, we consider a single-cell wireless edge system consisting of a finite set $\mathcal{D}$ of wireless devices,
a base station (BS) co-located with an edge server, and a remote cloud accessible through backhaul.
The devices in $\mathcal{D}$ may include mobile terminals, cameras, sensors, or other wireless endpoints involved in agent execution.
Multiple agents are deployed across different devices to support diverse applications, and each agent initiates a job when it needs to handle a task.
The time horizon is divided into equal-length slots, and $t\in\{0,1,2,\ldots\}$ denotes the slot index.
We consider a finite set $\mathcal{J}$ of jobs that arrive over time.
For each job $j$, let $a_j$ and $\omega_j$ denote its arrival slot and priority weight, respectively.

We assume a central scheduler at the BS to orchestrate agent job execution.
During each slot, newly generated workflow operations and their dependencies are reported to the scheduler.
The scheduler also obtains wireless channel information from the BS, monitors the available edge computing capacity,
and tracks the execution state of the workflow.
Based on this information, the scheduler jointly optimizes the communication and computation resources allocated to the agents.

\subsubsection{Dynamic workflow graph}
At the beginning of slot $t$, the revealed portion of job $j$ is represented by a dynamic workflow graph (DWG) $G_j(t)=\bigl(V_j(t),E_j(t)\bigr)$,
where $V_j(t)$ is the set of operations and $E_j(t)$ represents the precedence constraints among them.
We assume that $G_j(t)$ is a directed acyclic graph (DAG) for all $j$ and $t$.
The operation set is partitioned into disjoint computation and communication sets,
i.e., $V_j(t)=V_j^{\mathrm{c}}(t) \mathbin{\dot{\cup}} V_j^{\mathrm{m}}(t)$.

The workflow evolves as the agent generates new steps.
Let $\Delta V_j(t)$ and $\Delta E_j(t)$ denote the operations and precedence edges newly revealed during slot $t$.
Then,
\begin{equation*}
G_j(t+1)=\Bigl(V_j(t)\cup\Delta V_j(t),\;E_j(t)\cup\Delta E_j(t)\Bigr).
\end{equation*}
To preserve causal consistency during workflow revelation, the precedence constraints of
a newly revealed operation must be revealed with the operation itself:
\begin{property}[Causal Workflow Revelation]
\label{prop:causality}
For every job $j$ and slot $t$, we have
\begin{gather*}
\Delta V_j(t)\cap V_j(t)=\emptyset, \\
\Delta E_j(t)\subseteq \bigl(V_j(t)\cup\Delta V_j(t)\bigr)\times\Delta V_j(t), \\
(v,u)\in\Delta E_j(t) \Rightarrow v\in\Delta V_j(t) \ \text{or}\ v\text{ is a leaf node of }G_j(t).
\label{prop:batched_successors}
\end{gather*}
\end{property}
The first condition ensures that each operation is revealed only once.
The second condition requires every newly revealed edge to terminate at a newly revealed operation.
Therefore, all incoming edges of an operation are revealed together with the operation, and its predecessor set remains fixed thereafter.
The third condition requires every newly revealed edge to originate either from a newly revealed operation or from a leaf node of the current revealed graph,
which also implies that all immediate successors of an operation are revealed together.
For any operation $v$ of job $j$, let $a_v$ denote its revelation slot, $S_v$ and $F_v$ denote its start and finish slots, respectively.
An operation $v$ is ready if it has been revealed and all of its predecessors have finished. The corresponding ready slot is
\begin{equation*}
A_v\triangleq
\min\left\{t\ge a_v: t \ge F_u,\ \forall u\text{ such that }(u,v)\in E_j(a_v)\right\}.
\end{equation*}
To meet the precedence constraint, a feasible schedule must satisfy
\begin{equation}
  S_v\ge A_v, \quad \forall v\in V_j(t), \forall j\in\mathcal{J}, \forall t.
\label{eq:precedence_constraint}
\end{equation}

A job is completed when the agent planner terminates the workflow and all operations required for delivering the final response have completed.
Let $F_j$ denote the corresponding completion slot. The end-to-end latency of job $j$, measured in slots, is $C_j=F_j-a_j$.
The set of active jobs in slot $t$ is defined as
$\mathcal{J}(t)\triangleq\{j\in\mathcal{J}:a_j\le t<F_j\}$.

\subsubsection{Computation operations}
A computation operation $v_c\in V_j^{\mathrm{c}}(t)$ represents an agent execution step, such as preprocessing, model inference, aggregation, or tool execution.
Let $\mathcal{P}$ denote the set of operation types.
Each computation operation $v_c$ is associated with a type $p_{v_c}\in\mathcal{P}$ and an operation descriptor $\xi_{v_c}$,
which contains features relevant to its resource demand.
For example, $\xi_{v_c}$ may include the model identifier, input token length, image resolution, or tool category.

For each operation type $p\in\mathcal{P}$, the scheduler maintains a workload estimator $f_p$ based on historical data.
The initial workload of $v_c$ is estimated as $x_{v_c}^{\mathrm{c},0}=f_{p_{v_c}}(\xi_{v_c})$, measured in
floating-point operations (FLOPs).
Let $L_{v_c}\subseteq\{\mathrm{edge},\mathrm{cloud}\}\cup\mathcal{D}$ denote the feasible execution locations of operation $v_c$.
Note that $L_{v_c}$ may be a singleton when the corresponding operation can only be executed at a fixed location,
such as tool invocation or LLM inference in the cloud.
The operation placement $l_{v_c}\in L_{v_c}$ is selected once at the ready slot $A_{v_c}$ and remains unchanged thereafter.

\subsubsection{Communication operations}
A communication operation $v_m\in V_j^{\mathrm{m}}(t)$ represents the transfer of data required by the workflow.
Such an operation is introduced when an intermediate result, sensing input, tool request, or final response must be transmitted among devices and the edge server.
For brevity, this paper defines communication operations only for wireless transmission,
whereas wired transmission to the cloud is not represented as a communication operation and is instead modeled separately in Section~\ref{subsec:resource_model}.

For each communication operation $v_m$, let $x_{v_m}^{\mathrm{m},0}>0$ denote its initial workload,
i.e., the amount of data (measured in bits) to be transmitted.
The workload of some communication operations may not be available upon revelation.
For example, if an operation transmits the output of a tool invocation,
the data size may remain unknown until the preceding operation is completed.
In such cases, we estimate the transmission workload using the same approach as for computation operations
and replace the estimate with the actual value once the workload becomes available.
In addition, we use $e_{v_m}\in\mathcal{D}$ to denote the associated wireless endpoint,
and $h_{v_m}\in\mathcal{H}\triangleq\{\mathrm{UL},\mathrm{DL}\}$ to indicate whether the operation corresponds to uplink or downlink transmission.

\subsection{Wireless Edge Resource Model} \label{subsec:resource_model}
\subsubsection{Communication model}
For each direction $h\in\mathcal{H}$, let $N_h(t)$ denote the number of available wireless resource blocks (RBs) in slot $t$.
For each revealed communication operation $v_m$, let $n_{v_m}(t)\in\mathbb{Z}_{\ge0}$ denote the number of RBs allocated to it in slot $t$.
The RB allocation must satisfy the capacity constraint
\begin{equation}
\sum_{j\in\mathcal{J}(t)} \sum_{\substack{v_m\in V_j^{\mathrm{m}}(t):\;h_{v_m}=h}} n_{v_m}(t) \le N_h(t), \quad \forall h\in\mathcal{H},\;\forall t.
\label{eq:rb_constraint}
\end{equation}
For the associated endpoint $e_{v_m}$ and direction $h_{v_m}$,
we can obtain the spectral efficiency from channel quality indicator (CQI) feedback and the selected modulation and coding scheme (MCS).
Based on this spectral efficiency, we can calculate the amount of data that one RB can transmit for $v_m$ during slot $t$, denoted by $r_{v_m}(t)$.

Let $x_{v_m}^{\mathrm{m}}(t)$ denote the remaining workload of communication operation $v_m$ at slot $t$.
When $v_m$ is ready, it is initialized as $x_{v_m}^{\mathrm{m}}(A_{v_m})=x_{v_m}^{\mathrm{m},0}$
and evolves as
\begin{equation*}
x_{v_m}^{\mathrm{m}}(t+1)=\left[x_{v_m}^{\mathrm{m}}(t)-n_{v_m}(t)r_{v_m}(t)\right]^+, \quad t\ge A_{v_m},
\end{equation*}
where $[x]^+\triangleq\max\{x,0\}$.
Consequently, the start and finish slots of $v_m$ can be expressed as
\begin{equation*}
S_{v_m}=\min\{t\ge A_{v_m}:n_{v_m}(t)>0\},
\end{equation*}
and
\begin{equation*}
F_{v_m}=\min\{t\ge S_{v_m}:x_{v_m}^{\mathrm{m}}(t)=0\}.
\end{equation*}

\subsubsection{Computation model}
For an edge-executed computation operation $v_c$, let
$q_{v_c}(t)\ge0$ denote its allocated computing capacity in slot $t$,
i.e., the amount of workload that it can process during that slot.
Similar to the communication model, the allocation of edge computing resources must satisfy
\begin{equation}
\sum_{j\in\mathcal{J}(t)} \sum_{\substack{v_c\in V_j^{\mathrm{c}}(t):\;l_{v_c}=\mathrm{edge}}} q_{v_c}(t) \le Q_{\mathrm{e}}(t), \quad \forall t,
\label{eq:edge_constraint}
\end{equation}
where $Q_{\mathrm{e}}(t)$ is the total workload that the edge server can process in slot $t$.
Let $x_{v_c}^{\mathrm{c}}(t)$ denote the remaining workload of $v_c$ at the beginning of slot $t$.
When an edge-placed operation $v_c$ becomes ready, its remaining workload is initialized as $x_{v_c}^{\mathrm{c}}(A_{v_c})=x_{v_c}^{\mathrm{c},0}$
and evolves as
\begin{equation*}
x_{v_c}^{\mathrm{c}}(t+1)=\left[x_{v_c}^{\mathrm{c}}(t)-q_{v_c}(t)\right]^+,
\quad t\ge A_{v_c}.
\end{equation*}
As a result, for an edge-executed computation operation, the start and finish slots are
\begin{equation*}
S_{v_c}=\min\{t\ge A_{v_c}:q_{v_c}(t)>0\},
\end{equation*}
and
\begin{equation*}
F_{v_c}=\min\{t\ge S_{v_c}:x_{v_c}^{\mathrm{c}}(t)=0\}.
\end{equation*}

For a cloud-executed computation operation, the backhaul transmission and cloud computing capacities are beyond the control of the BS-side
scheduler and are estimated from historical data.
Specifically, the transmission delay incurred by cloud execution is modeled as
\begin{equation*}
\tau_{v_c}^{\mathrm{cloud}} = 2 \tau^{\mathrm{fix}}+\frac{s_{v_c}^{\mathrm{in}}}{R^{\mathrm{UL}}} +\frac{s_{v_c}^{\mathrm{out}}}{R^{\mathrm{DL}}},
\end{equation*}
where $s_{v_c}^{\mathrm{in}}$ and $s_{v_c}^{\mathrm{out}}$ are the input and output data sizes for cloud execution,
$R^{\mathrm{UL}}$ and $R^{\mathrm{DL}}$ denote the profiled amounts of uplink and downlink data transmitted per slot,
and $\tau^{\mathrm{fix}}$ is the fixed one-way propagation and processing delay.
Let $\bar q_{v_c}$ be the profiled cloud computing capacity for operation $v_c$,
then the duration of a cloud-executed computation operation is
\begin{equation*}
T_{v_c}^{\mathrm{cloud}}
=\left\lceil
\tau_{v_c}^{\mathrm{cloud}}+\frac{x_{v_c}^{\mathrm{c},0}}{\bar q_{v_c}}
\right\rceil.
\end{equation*}
If $v_c$ is placed at the cloud, it is submitted immediately when ready. Hence, its start and finish slots are
\begin{equation*}
S_{v_c}=A_{v_c},
\quad
F_{v_c}=S_{v_c}+ T_{v_c}^{\mathrm{cloud}}.
\end{equation*}

Some computation operations (e.g., tool invocations) are executed at wireless devices.
For such an operation $v_c$, the feasible location set is a singleton
$L_{v_c}=\{e_{v_c}\}$, where $e_{v_c}\in\mathcal{D}$ is the device that provides the corresponding service.
Let $T_{v_c}^{\mathrm{dev}}$ denote the profiled execution time (measured in slots) of the service.
Then the start and finish slots of $v_c$ are
\begin{equation*}
S_{v_c}=A_{v_c},
\quad
F_{v_c}=S_{v_c}+T_{v_c}^{\mathrm{dev}}.
\end{equation*}

\subsection{Online Workflow-Aware Scheduling Problem} \label{subsec:problem_formulation}
To facilitate the execution of agent jobs, the scheduler
observes the current information state at the beginning of each slot $t$, denoted by
\begin{equation*}
\mathcal{I}(t)=\left\{ G_j(t),\mathcal{O}_j(t),\omega_j, N_h(t),Q_{\mathrm{e}}(t)\right\}_{j\in\mathcal{J}(t), h\in\mathcal{H}}
\end{equation*}
where $\mathcal{O}_j(t)$ denotes the operation states of job $j$,
including whether each revealed operation is blocked, ready, or finished,
as well as its remaining workloads $x_{v_c}^{\mathrm{c}}(t)$ and $x_{v_m}^{\mathrm{m}}(t)$ 
for computation and communication operations, respectively, and per-RB transmission capacity $r_{v_m}(t)$
for communication operations, and placement decision if already made.
Let $\bm{a}(t)$ collect all placement and resource allocation decisions made in slot $t$, and define the information history available to the scheduler as
\begin{equation*}
\mathcal{H}_t\triangleq
\bigl(\mathcal{I}(0),\bm{a}(0),\ldots,
\mathcal{I}(t-1),\bm{a}(t-1),\mathcal{I}(t)\bigr).
\end{equation*}
A causal scheduling policy $\pi=\{\pi_t\}_{t\ge0}$ maps this history to
\begin{equation*}
\pi_t(\mathcal{H}_t)=\bm{a}(t)
=\left(
\{l_{v_c}\}_{v_c\in\mathcal{U}(t)},
\{q_{v_c}(t)\},
\{n_{v_m}(t)\}
\right),
\end{equation*}
where $\mathcal{U}(t)$ denotes the set of computation operations that first become ready in slot $t$.
Note that decisions in slot $t$ can depend only on $\mathcal{H}_t$.
Let $\Pi$ denote the set of causal policies that satisfy all feasibility constraints and complete every
job almost surely with finite expected weighted latency.
Then the online workflow-aware scheduling problem is formulated as
\begin{subequations}
\label{prob:online_scheduling}
\begin{align}
  \min_{\pi\in\Pi}\quad & \mathbb{E}_{\pi}\left[\sum_{j\in\mathcal{J}}\omega_j C_j\right] \label{prob:objective}\\
  \text{s.t.}\quad & \eqref{eq:precedence_constraint}, \eqref{eq:rb_constraint}, \eqref{eq:edge_constraint} \label{prob:precedence}\\
                 & l_{v_c}\in L_{v_c}, \quad \forall v_c\in\mathcal{U}(t),\forall t \label{prob:placement}\\
                 & q_{v_c}(t) \geq 0, \quad \forall v_c\in V_j^{\mathrm{c}}(t),\forall j,t \notag\\
                 & n_{v_m}(t)\in\mathbb{Z}_{\ge0}, \quad \forall v_m\in V_j^{\mathrm{m}}(t),\forall j,t \label{prob:rb_integer}
\end{align}
\end{subequations}
where the expectation is taken over workflow evolution, wireless channel variations, and fluctuations in resource availability.
Constraint~\eqref{prob:precedence} collects the precedence, wireless RB allocation, and edge computing capacity constraints introduced above.
Constraints~\eqref{prob:placement}-\eqref{prob:rb_integer} specify the feasible domains of the control variables.

This problem is challenging for three reasons.
First, the workflow is dynamically generated, and the scheduler must make decisions without knowing the complete dependency graph.
Second, communication and computation are tightly coupled through workflow dependencies: wireless transmissions determine when downstream computation operations become ready,
while computation placement determines their processing and transmission delays.
Third, wireless channel states and edge computing availability vary over time, limiting the effectiveness of static scheduling.
These challenges motivate an online scheduler that accounts for workflow dependencies and resource dynamics jointly.

\section{Algorithm Design} \label{sec:algorithm}
The main difficulty in solving \eqref{prob:online_scheduling} is that the completion time of a job depends on future workflow operations, 
resource availability, and channel conditions, which are unknown when the current scheduling decision is made.
To address this issue, we first transform the weighted completion time objective into an equivalent per-slot holding-cost objective.
Based on this reformulation, the optimal online policy can be characterized by a Bellman equation.
However, directly solving the Bellman equation is computationally impractical because the future workflow and resource states are unknown.
We therefore construct a tractable cost-to-go surrogate based on the revealed workflow and the estimated remaining latency of each active job.
We use this surrogate to compare computation-placement actions and derive the criticality of each ready operation 
for shared wireless and edge-computing resource allocation.

\subsection{Objective Reformulation}
The objective in \eqref{prob:objective} is determined by the completion slot $F_j$ of each job.
However, when the scheduler makes a decision in slot $t$, the value of $F_j$ is unknown, 
as it depends on unrevealed workflow operations, future resource availability, channel conditions, and subsequent scheduling decisions.
Therefore, the original completion time objective is difficult to optimize directly in an online setting.

To address this issue, we transform the job completion latency into a per-slot holding cost. For job $j$, its latency can be expressed as
\begin{equation}
C_j=F_j-a_j =\sum_{t=0}^{\infty} \mathbf{1}\{a_j\le t<F_j\}.
\label{eq:job_holding_cost}
\end{equation}
Multiplying \eqref{eq:job_holding_cost} by $\omega_j$, summing over all jobs, and exchanging the order of summation yield
\begin{equation}
\sum_{j\in\mathcal{J}}\omega_j C_j =\sum_{t=0}^{\infty} \sum_{j\in\mathcal{J}(t)}\omega_j.
\label{eq:holding_cost_identity}
\end{equation}
Equation~\eqref{eq:holding_cost_identity} shows that the weighted completion time objective is equivalent to the cumulative holding cost of all active jobs. 
Accordingly, we define the holding cost in slot $t$ as $c(t)\triangleq \sum_{j\in\mathcal{J}(t)}\omega_j$.
Given the history $\mathcal{H}_t$, the optimal expected remaining cost is defined as
\begin{equation*}
J^*(\mathcal{H}_t)\triangleq \min_{\pi\in\Pi} \mathbb{E}_{\pi}\!\left[ \sum_{\tau=t}^{\infty}c(\tau)\,\middle|\,\mathcal{H}_t \right].
\end{equation*}
The quantity $J^*(\mathcal{H}_t)$ is referred to as the optimal cost-to-go and satisfies the following Bellman equation:
\begin{equation}
J^*(\mathcal{H}_t)=c(t)+
\min_{\bm{a}(t)\in\mathcal{A}(\mathcal{I}(t))}
\mathbb{E}\!\left[J^*(\mathcal{H}_{t+1})
\mid\mathcal{H}_t,\bm{a}(t)\right],
\label{eq:history_bellman}
\end{equation}
where $\mathcal{A}(\mathcal{I}(t))$ denotes the set of feasible scheduling decisions satisfying constraints 
\eqref{prob:precedence}-\eqref{prob:rb_integer} in slot $t$.

Theoretically, the optimal scheduling decisions can be obtained by directly solving the Bellman equation \eqref{eq:history_bellman}.
However, this is computationally impractical because the next state depends on unrevealed workflow operations and future resource states. 
In the following subsection, we construct a tractable cost-to-go surrogate $\widehat J(\mathcal H_t)$ based on the 
revealed workflow and the estimated remaining latencies of active jobs. 

\subsection{Cost-to-Go Surrogate} \label{subsec:latency_cost_to_go}
In this subsection, we first convert the remaining communication and computation workloads into estimated durations, 
so that heterogeneous operation types can be compared under a unified metric. 
After that, we can further estimate the remaining latencies of workflow branches and active jobs based on the revealed DWG.
These estimates are finally aggregated into the look-ahead cost-to-go surrogate $\widehat J(\mathcal{H}_t)$.

\subsubsection{Converting Workload to Estimated Duration}
In the DWG, communication and computation operations are characterized by different workload units:
communication workloads are measured in bits, whereas computation workloads are measured in FLOPs.
These heterogeneous quantities cannot be directly compared or accumulated along a workflow path.
We therefore convert both types of workloads into estimated durations.

For an active job $j$, let $\widetilde V_j(t)$ denote the set of revealed but unfinished operations. 
For each $v\in\widetilde V_j(t)$, let $\widehat d_v(t)$ denote its estimated remaining duration, measured in slots. 
Let $\overline\mu_{v_m}^{\mathrm{m}}(t)$ and $\overline\mu_{v_c}^{\mathrm{c}}(t)$ denote the reference communication and computation service rates, 
respectively, estimated from historical data and the current system state. 
In our implementation, the reference service rate is obtained by evenly dividing the currently available resources 
among ready operations and smoothing the resulting per-operation rate over time. 
Then we have $\widehat d_{v_m}(t)=x_{v_m}^{\mathrm{m}}(t)/\overline\mu_{v_m}^{\mathrm{m}}(t)$ for communication operations and
$\widehat d_{v_c}(t)=x_{v_c}^{\mathrm{c}}(t)/\overline\mu_{v_c}^{\mathrm{c}}(t)$ for computation operations.
It is worth noting that $\widehat d_v(t)$ is not intended to predict the exact completion time of operation $v$,
since future resource sharing and channel conditions remain unknown.
Instead, its purpose is to express all operations in the same time unit so that their workflow branches can be compared.

\subsubsection{Remaining Latency of Branches}
The duration of an operation alone does not fully characterize its scheduling criticality. 
For example, a short operation may be followed by a long sequence of successors, 
whereas a longer operation on a short parallel branch may have little impact on the overall job completion time. 
We therefore estimate the remaining latency of each workflow branch by accumulating the estimated durations along the revealed DWG.

For each $v\in\widetilde V_j(t)$, let $\widehat L_v(t)$ denote the estimated remaining branch latency starting from operation $v$, including its own duration. 
Let $\mathcal{S}_v(t)$ denote the set of revealed but unfinished immediate successors of $v$. 
If $\mathcal{S}_v(t)=\emptyset$, $v$ is a current leaf operation. 
Its completion may reveal additional downstream operations, 
whose estimated latency is denoted by $\widehat L_v^{\mathrm{tail}}(t)$ and is obtained from the observed history. 
By Property~\ref{prop:causality}, an operation with revealed successors cannot acquire additional successors later. 
Therefore, tail-latency prediction is required only for current leaf operations. 
We set $\widehat L_v^{\mathrm{tail}}(t)=0$ when the agent declares that the branch has terminated or when tail prediction is disabled.

For a non-leaf operation, the downstream latency is determined by the longest successor branch,
expressed as $\max\{\widehat L_u(t):u\in\mathcal{S}_v(t)\}$. 
However, using the ordinary maximum introduces two issues.
First, the $\max$ operator is non-differentiable, which complicates the definition of operation criticality described later. 
Second, it assigns all criticality to the currently longest successor and ignores branches with similar remaining latencies,
which may become the new bottleneck after a small amount of workload progress. 
To overcome this problem, we instead use the following smooth approximation
\begin{equation*}
\operatorname{smax}_{\kappa}(\{z_i\}) \triangleq \frac{1}{\kappa}\log\!\left(\sum_i e^{\kappa z_i}\right),
\end{equation*}
where $\kappa$ is a tunable parameter. 
A larger $\kappa$ makes $\operatorname{smax}_{\kappa}$ closer to the ordinary maximum, 
whereas a smaller $\kappa$ distributes the branch sensitivity more evenly across different successors.

Combining the leaf and non-leaf cases, the estimated remaining branch latency is given by
\begin{equation}
\widehat L_v(t)=\widehat d_v(t)+
\begin{cases}
\operatorname{smax}_{\kappa} \bigl(\{\widehat L_u(t)\}_{u\in\mathcal{S}_v(t)}\bigr), &\mathcal{S}_v(t)\ne\emptyset,\\
\widehat L_v^{\mathrm{tail}}(t),&\mathcal{S}_v(t)=\emptyset
\end{cases}.
\label{eq:remaining_branch_latency}
\end{equation}

\subsubsection{Look-Ahead Cost-to-Go Surrogate}
While $\widehat L_v(t)$ characterizes the estimated remaining latency of an individual branch, 
the scheduling objective depends on the end-to-end latency of each job. 
We therefore aggregate the branch-level estimates over the ready operations of job $j$ and define its estimated remaining latency as
\begin{equation}
\widehat T_j(t)=\operatorname{smax}_{\kappa} \bigl(\{\widehat L_v(t):v\in\mathcal{R}_j(t)\}\bigr),
\label{eq:job_remaining_latency}
\end{equation}
where $\mathcal{R}_j(t)\subseteq\widetilde V_j(t)$ denotes the set of ready operations.

Since an unfinished job $j$ incurs a holding cost of $\omega_j$ in each slot, 
its remaining holding cost can be directly estimated as $\omega_j\lceil\widehat T_j(t)\rceil$. 
However, this estimate treats jobs independently and thus cannot capture the coupling among job completion times induced by resource sharing and waiting. 
For example, consider two equal-weight jobs with estimated remaining latencies of one and twenty slots, respectively. 
Serving the first job completes it and reduces the next-slot holding cost from $2\omega$ to $\omega$, 
whereas serving the second job leaves both jobs active and keeps the next-slot holding cost at $2\omega$. 
Nevertheless, the linear surrogate evaluates the two processing orders equivalently, 
as the total estimated remaining latency is twenty slots in both cases.

The preceding example shows that the linear surrogate depends only on the total estimated remaining latency 
and cannot distinguish between completing a nearly finished job and making the same amount of progress on a long job. 
To reflect this difference, we assign decreasing weights to holding costs incurred farther in the estimated future. 
Specifically, for an offset $\tau\geq0$, we define $h_{\zeta}(\tau)\triangleq\frac{\zeta}{\zeta+\tau}$, 
where $\zeta>0$ is a tunable parameter. 
Applying this weight to the future holding costs of all active jobs yields the discrete look-ahead cost-to-go surrogate
\begin{equation}
\widehat J'(\mathcal H_t)
\triangleq
\sum_{j\in\mathcal J(t)}\omega_j
\sum_{\tau=0}^{\lceil\widehat T_j(t)\rceil-1}
h_{\zeta}(\tau).
\label{eq:discrete_look_ahead_cost}
\end{equation}
To preserve differentiability, we replace the discrete sum with the following continuous relaxation:
\begin{equation*}
\sum_{\tau=0}^{\lceil T\rceil-1} \frac{\zeta}{\zeta+\tau} \approx g_{\zeta}(T) \triangleq
\int_0^T \frac{\zeta}{\zeta+\tau}\,d\tau =\zeta\log\!\left(1+\frac{T}{\zeta}\right).
\end{equation*}
The resulting continuous look-ahead cost-to-go surrogate is\footnote{
  This formulation accounts solely for currently active jobs and omits future arrivals,
  as only active jobs have observable workflow states and remaining workloads at slot $t$.
}
\begin{equation}
\widehat J(\mathcal{H}_t) \triangleq \sum_{j\in\mathcal{J}(t)} \omega_j g_{\zeta}\!\left(\widehat T_j(t)\right).
\label{eq:cost_to_go_surrogate}
\end{equation}

\subsection{Operation Criticality} \label{subsec:operation_criticality}
This subsection derives the operation criticality used to allocate shared wireless and edge-computing resources. 
We first characterize the marginal cost reduction obtained by processing an additional unit of workload before any ready operation completes. 
We then incorporate the additional cost reduction associated with operation completion.

\subsubsection{Marginal Cost Reduction}
Having constructed the look-ahead cost-to-go surrogate $\widehat J(\mathcal H_t)$, 
we next quantify the cost reduction produced by processing an additional unit of workload. 
If no ready operation completes, processing operation $v$ only reduces its remaining workload. 
The resulting reduction in the surrogate cost can therefore be characterized by its derivative with respect to $x_v(t)$.

For a ready operation $v\in\mathcal{R}_j(t)$, let $\varphi_{jv}(t)$ denote its coefficient in the smoothed maximum in \eqref{eq:job_remaining_latency}, given by
\begin{equation*}
\varphi_{jv}(t)
\triangleq\frac{e^{\kappa\widehat L_v(t)}} {\sum_{u\in\mathcal{R}_j(t)}e^{\kappa\widehat L_u(t)}}.
\end{equation*}
Applying the chain rule to the cost-to-go surrogate, the smoothed maximum, and the workload-to-duration conversion 
gives the marginal cost reductions for communication and computation operations:
\begin{equation}
\begin{aligned}
\beta_{v_m}(t)
&\triangleq \frac{\partial\widehat J(\mathcal{H}_t)}
{\partial x_{v_m}^{\mathrm{m}}(t)}
=\frac{\omega_j\zeta}{\zeta+\widehat T_j(t)} \times
\frac{\varphi_{jv_m}(t)}{\overline\mu_{v_m}^{\mathrm{m}}(t)},\\
\beta_{v_c}(t)
&\triangleq \frac{\partial\widehat J(\mathcal{H}_t)}
{\partial x_{v_c}^{\mathrm{c}}(t)}
=\frac{\omega_j\zeta}{\zeta+\widehat T_j(t)} \times
\frac{\varphi_{jv_c}(t)}{\overline\mu_{v_c}^{\mathrm{c}}(t)}.
\end{aligned}
\label{eq:operation_criticality}
\end{equation}
For convenience, we use $\beta_v(t)$ to uniformly represent $\beta_{v_m}(t)$ and $\beta_{v_c}(t)$.
By definition, $\beta_v(t)$ gives the reduction in surrogate cost obtained by processing one additional unit of workload of operation $v$. 
The first term in the right-hand side of \eqref{eq:operation_criticality} is larger for jobs with higher weights and shorter estimated remaining latencies. 
The second term is large when branch $v$ is more likely to determine the job latency,
while the second term converts one unit of workload into its estimated processing duration under the reference service rate.
However, $\beta_v(t)$ is derived under the assumption that operation $v$ remains unfinished. 
Hence, it does not capture the discrete change in the revealed workflow when $v$ completes. 
The next subsection accounts for this additional cost reduction.

\subsubsection{Cost Reduction from Operation Completion}
The marginal cost reduction $\beta_v(t)$ only describes the effect of processing an unfinished operation. 
Completing operation $v$ may additionally make its revealed successors ready and change the reference service rates of other ready operations. 
To capture this additional reduction, we construct a hypothetical state $\widehat{\mathcal H}_t^{v,+}$ 
for each ready communication or computation operation $v$ with $x_v(t)>0$, in which $v$ has just completed. 
We first mark $v$ as finished and set its remaining workload to zero. 
We then add each revealed successor whose predecessors have all finished to the set of ready operations. 
If $v$ is a revealed leaf and workflow termination has not been declared, 
we retain its predicted continuation $\widehat L_v^{\mathrm{tail}}(t)$ as a virtual downstream operation. 
Finally, we recompute the reference communication and edge service rates for the resulting active operations. 

Let $\widehat J^{v,+}(\mathcal H_t)$ denote the cost-to-go surrogate in this hypothetical state. 
The reduction caused by completing $v$ is defined as
\begin{gather}
\widehat J^{v,+}(\mathcal H_t) \triangleq \widehat J(\widehat{\mathcal H}_t^{v,+}), \label{eq:virtual_completion_cost}\\
D_v(t) \triangleq \left[\widehat J(\mathcal H_t) -\widehat J^{v,+}(\mathcal H_t)\right]^+.\notag
\end{gather}
To compare this finite reduction with $\beta_v(t)$, which is defined per unit of workload, we distribute $D_v(t)$ over the remaining workload of $v$. 
Hence, we have
\begin{equation}
\beta_v^{+}(t)
\triangleq\frac{D_v(t)}{x_v(t)},
\qquad x_v(t)>0.
\label{eq:completion_gain_per_unit}
\end{equation}
Unlike $\beta_v(t)$, $\beta_v^{+}(t)$ captures the cost reduction caused by the workflow changes at completion.

We define the operation criticality as the larger of the continuous marginal cost reduction and the average cost reduction at completion:
\begin{equation}
\widetilde\beta_v(t) \triangleq \max\!\left\{\beta_v(t),\beta_v^{+}(t)\right\}.
\label{eq:completion_corrected_criticality}
\end{equation}
To illustrate the cost reduction from completing an operation, 
consider an isolated job with one ready terminal operation and estimated duration $\widehat d_v(t)=x_v(t)/\overline\mu_v(t)$. 
Completing the operation leaves no remaining cost, and the concavity of $g_{\zeta}$ gives $g_{\zeta}(\widehat d_v)\ge g_{\zeta}'(\widehat d_v)\widehat d_v$. Hence,
\begin{equation*}
\widetilde\beta_v(t)=\omega_j g_{\zeta}(\widehat d_v(t))/x_v(t).
\end{equation*}
Because $g_{\zeta}(d)/d$ decreases with $d$, completing a shorter weighted operation gives a larger average cost reduction. For a general workflow, $D_v(t)$ is computed from the hypothetical post-completion state, which includes newly ready revealed successors and, when applicable, the predicted continuation of the workflow. The following subsections use the full cost-to-go surrogate for computation placement and $\widetilde\beta_v(t)$ for shared-resource allocation.

\subsection{Computation Placement} \label{subsec:placement}
The execution location of a computation operation affects not only its own completion time but also the completion times of other active operations. 
In particular, placing an operation at the edge may increase the workload of the shared edge server and delay other edge-executed operations. 
Therefore, computation placement is determined by its effect on the cost-to-go surrogate of all active jobs, rather than by the execution time of the current operation alone.

For each newly ready computation operation $v_c\in\mathcal{U}(t)$, the scheduler evaluates every feasible location $l\in L_{v_c}$. 
The estimated duration of executing $v_c$ at location $l$ is
\begin{equation*}
\widehat d_{v_c}(t;l)=
\begin{cases}
x_{v_c}^{\mathrm{c},0}/\overline\mu_{v_c}^{\mathrm{c}}(t;l),&l=\mathrm{edge},\\
T_{v_c}^{\mathrm{cloud}},&l=\mathrm{cloud},\\
T_{v_c}^{\mathrm{dev}},&l\in\mathcal{D}.
\end{cases},
\end{equation*}
Here, $\overline\mu_{v_c}^{\mathrm{c}}(t;l)$ denotes the reference service rate of $v_c$ when it is placed at location $l$. 
If $l=\mathrm{edge}$, the scheduler updates the reference service rates of all edge-executed operations to account for the resulting change in edge workload. It then propagates the updated durations backward through the revealed DWG to obtain the candidate branch latencies $\widehat L_u(t;l_{v_c}=l)$ and the corresponding job latencies
\begin{equation*}
\widehat T_j(t;l_{v_c}=l)
=\operatorname{smax}_{\kappa}
\bigl(\{\widehat L_u(t;l_{v_c}=l):u\in\mathcal{R}_j(t)\}\bigr).
\end{equation*}
The resulting cost-to-go surrogate under location $l$ is
\begin{equation}
\widehat J(\mathcal{H}_t;l_{v_c}=l)
\triangleq
\sum_{j\in\mathcal{J}(t)}
\omega_j g_{\zeta}\!\left(\widehat T_j(t;l_{v_c}=l)\right).
\label{eq:placement_cost}
\end{equation}
The scheduler selects the feasible location with the smallest surrogate cost:
\begin{equation}
l_{v_c}^{*}\in\argmin_{l\in L_{v_c}} \widehat J(\mathcal{H}_t;l_{v_c}=l).
\label{eq:placement_decision}
\end{equation}
When only a few computation operations become ready simultaneously, the scheduler evaluates all feasible joint placement combinations and selects the one with the smallest surrogate cost. When exhaustive evaluation is prohibitively expensive, it instead processes the operations sequentially in descending order of their job weights. After selecting the location of each operation, the scheduler updates the candidate edge workload and the associated reference service rates before evaluating the next operation.

\subsection{Resource Allocation} \label{subsec:resource_allocation}
After the computation locations have been selected, the scheduler allocates wireless RBs and edge computing capacity in the current slot. 
The selected locations determine the active edge workload and the corresponding reference service rates. 
Specifically, the communication and computation operations eligible for resource allocation in slot $t$ are
\begin{align*}
\mathcal{R}^{\mathrm{m}}_h(t)
&\triangleq\bigcup_{j\in\mathcal{J}(t)}
\bigl\{v_m\in\mathcal{R}_j(t)\cap V_j^{\mathrm{m}}(t):\\[-0.6ex]
&\hspace{7em}x_{v_m}^{\mathrm{m}}(t)>0,\ h_{v_m}=h\bigr\},\\
\mathcal{R}^{\mathrm{c}}_{\mathrm{e}}(t)
&\triangleq\bigcup_{j\in\mathcal{J}(t)}
\bigl\{v_c\in\mathcal{R}_j(t)\cap V_j^{\mathrm{c}}(t):\\[-0.6ex]
&\hspace{5.3em}x_{v_c}^{\mathrm{c}}(t)>0,\ l_{v_c}=\mathrm{edge}\bigr\}.
\end{align*}
The set $\mathcal{R}^{\mathrm{m}}_h(t)$ contains the ready communication operations in direction $h$, whereas $\mathcal{R}^{\mathrm{c}}_{\mathrm{e}}(t)$ contains the ready computation operations placed at the edge. For a given allocation, their completed workloads in slot $t$ are
\begin{gather*}
y_{v_m}^{\mathrm{m}}(t) \triangleq\min\{x_{v_m}^{\mathrm{m}}(t),n_{v_m}(t)r_{v_m}(t)\},\\
y_{v_c}^{\mathrm{c}}(t) \triangleq\min\{x_{v_c}^{\mathrm{c}}(t),q_{v_c}(t)\}.
\end{gather*}

Recall that $\widetilde\beta_v(t)$ is the estimated cost reduction per unit of completed workload. 
Then the optimal resource allocation can be obtained by solving
\begin{equation}
\begin{aligned}
\max_{\{n_{v_m}(t),q_{v_c}(t)\}}\quad
&\sum_{h\in\mathcal{H}}\sum_{v_m\in\mathcal{R}^{\mathrm{m}}_h(t)}
\widetilde\beta_{v_m}(t)y_{v_m}^{\mathrm{m}}(t)\\[-0.3ex]
&+\sum_{v_c\in\mathcal{R}^{\mathrm{c}}_{\mathrm{e}}(t)}
\widetilde\beta_{v_c}(t)y_{v_c}^{\mathrm{c}}(t)\\
\mathrm{s.t.}\quad
&\sum_{v_m\in\mathcal{R}^{\mathrm{m}}_h(t)}n_{v_m}(t)
\le N_h(t),\quad \forall h,\\
&\sum_{v_c\in\mathcal{R}^{\mathrm{c}}_{\mathrm{e}}(t)}q_{v_c}(t)
\le Q_{\mathrm{e}}(t),\\
&n_{v_m}(t)\in\mathbb{Z}_{\ge0},\quad q_{v_c}(t)\ge0.
\end{aligned}
\label{prob:slot_allocation}
\end{equation}
The objective and constraints in \eqref{prob:slot_allocation} are separable and piecewise linear.
Therefore, the resource allocation for communication and computation operations can be solved independently.
For RB allocation, suppose that $v_m$ has already received $k-1$ RBs in slot $t$. Assigning one additional RB to $v_m$ yields the marginal surrogate-cost reduction
\begin{equation}
\widetilde\beta_{v_m}(t) \min\!\bigl\{r_{v_m}(t),[x_{v_m}^{\mathrm{m}}(t)-(k-1)r_{v_m}(t)]^+\bigr\}.
\label{eq:radio_index}
\end{equation}
This quantity is nonincreasing in $k$. Hence, the radio subproblem is solved by repeatedly assigning one RB to the operation with the largest positive marginal reduction.

For edge computing resource allocation, the objective is linear in $q_{v_c}(t)$.
Consequently, we can sort the ready edge-executed operations in descending order of $\widetilde\beta_{v_c}(t)$ and iteratively allocate the remaining edge capacity.
Thus, we have
\begin{equation}
q_{v_c}(t)=\min\left\{ x_{v_c}^{\mathrm{c}}(t),Q_{\mathrm{e}}^{\mathrm{rem}}(t) \right\},
\label{eq:edge_greedy_fill}
\end{equation}
where $Q_{\mathrm{e}}^{\mathrm{rem}}(t)$ is the amount of remaining edge computing capacity before serving $v_c$.

\begin{algorithm}[t]
\caption{Online Scheduling Procedure}
\label{alg:online_procedure}
\begin{algorithmic}[1]
\FOR{each slot $t=0,1,\ldots$}
\STATE Observe $\mathcal{I}(t)$ and append it to $\mathcal{H}_t$.
\STATE Update the unfinished revealed graph of every active job and identify $\mathcal{U}(t)$.
\STATE Estimate the reference per-slot service rates and $\widehat L_v^{\mathrm{tail}}(t)$ from the observed history.
\STATE Evaluate the candidate $\widehat J(\mathcal{H}_t;l_{v_c}=l)$ surrogate costs, select the locations of operations in $\mathcal{U}(t)$ using \eqref{eq:placement_decision}, and update the estimated edge workload.
\STATE Compute $\widehat d_v(t)$, $\widehat L_v(t)$, $\widehat T_j(t)$, $\widehat J(\mathcal{H}_t)$, and $\beta_v(t)$ using \eqref{eq:remaining_branch_latency}--\eqref{eq:operation_criticality}.
\STATE For each ready shared-resource operation $v$, construct $\widehat{\mathcal H}_t^{v,+}$ and compute $D_v(t)$, $\beta_v^{+}(t)$, and $\widetilde\beta_v(t)$ using \eqref{eq:virtual_completion_cost}--\eqref{eq:completion_corrected_criticality}.
\FOR{each direction $h\in\mathcal{H}$}
\STATE Allocate up to $N_h(t)$ RBs one at a time using \eqref{eq:radio_index}.
\ENDFOR
\STATE Allocate edge computing resources in descending order of $\widetilde\beta_{v_c}(t)$ using \eqref{eq:edge_greedy_fill}.
\STATE Execute the selected actions and record $\bm{a}(t)$.
\ENDFOR
\end{algorithmic}
\end{algorithm}

Algorithm~\ref{alg:online_procedure} summarizes the proposed scheduler. 
Let $M(t)\triangleq\sum_{j\in\mathcal{J}(t)}(|\widetilde V_j(t)|+|E_j(t)|)$ denote the size of the active revealed graphs, and let $K(t)$ denote the number of ready operations that can receive shared communication or edge resources. 
Computing the branch latencies and marginal cost reductions requires one graph traversal and takes $O(M(t))$ time. 
A direct implementation then recomputes the surrogate for each hypothetical completion, which takes $O(K(t)M(t))$ time to compute all values of $D_v(t)$. 
This cost can be reduced by caching unaffected branch latencies and updating only the completed operation's job and the operations whose reference service rates change.

For resource allocation, the implementation of RB allocation takes 
$O(\sum_{h\in\mathcal{H}}(|\mathcal{R}^{\mathrm{m}}_h(t)|+N_h(t))\log |\mathcal{R}^{\mathrm{m}}_h(t)|)$ time, 
and sorting the ready edge operations takes $O(|\mathcal{R}^{\mathrm{c}}_{\mathrm{e}}(t)|\log |\mathcal{R}^{\mathrm{c}}_{\mathrm{e}}(t)|)$ time. 
Finally, let $P(t)=\prod_{v\in\mathcal U(t)}|L_v|$ denote the number of joint placement combinations. Exhaustive placement takes $O(P(t)M(t))$ time when $P(t)$ is below the configured threshold. Otherwise, the sequential fallback takes $O(\sum_{v\in\mathcal U(t)}|L_v|M(t))$ time. The placement step can also reuse branch-latency calculations that remain unchanged across candidate locations.

\section{Performance Evaluation}

\begin{figure*}[t]
\centering
\setlength{\abovecaptionskip}{0pt}
\begin{minipage}[t]{0.318\textwidth}
\vspace{0pt}
\centering
\includegraphics[width=\linewidth]{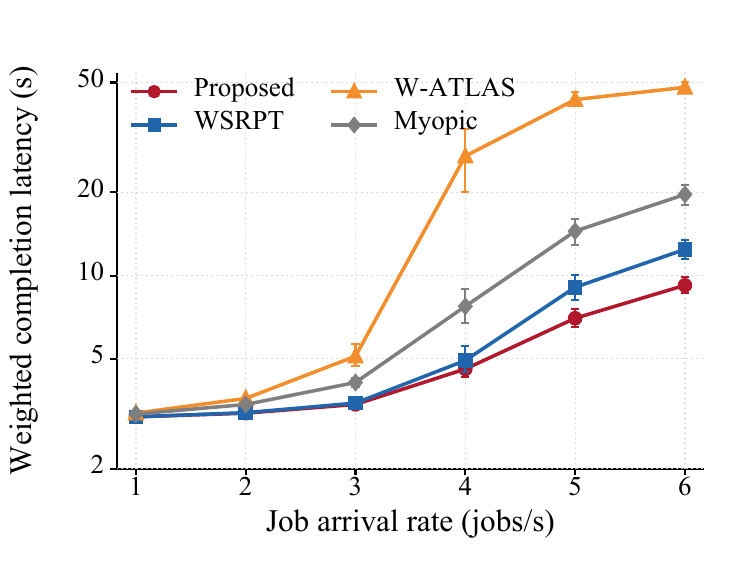}
\caption{Weighted mean completion latency versus job arrival rate.}
\label{fig:load_performance}
\end{minipage}
\hfill
\begin{minipage}[t]{0.318\textwidth}
\vspace{0pt}
\centering
\includegraphics[width=\linewidth]{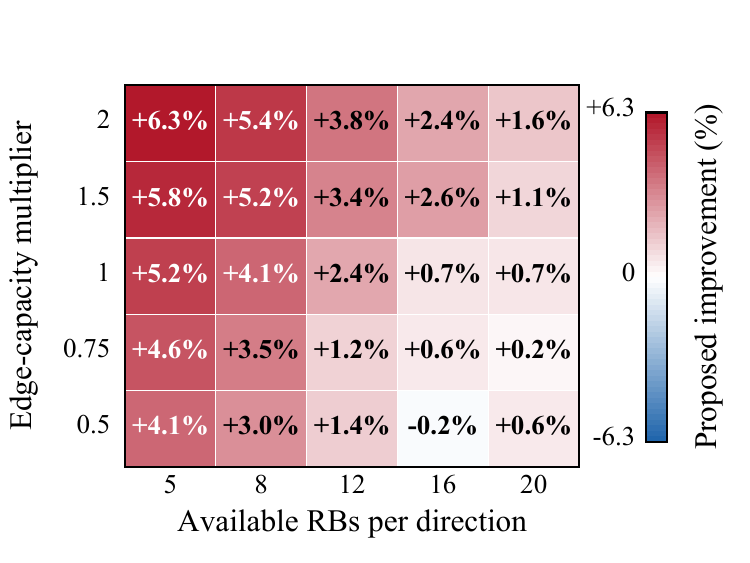}
\caption{Improvement over benchmarks under different resource levels.}
\label{fig:resource_regime}
\end{minipage}
\hfill
\begin{minipage}[t]{0.318\textwidth}
\vspace{0pt}
\centering
\includegraphics[width=\linewidth]{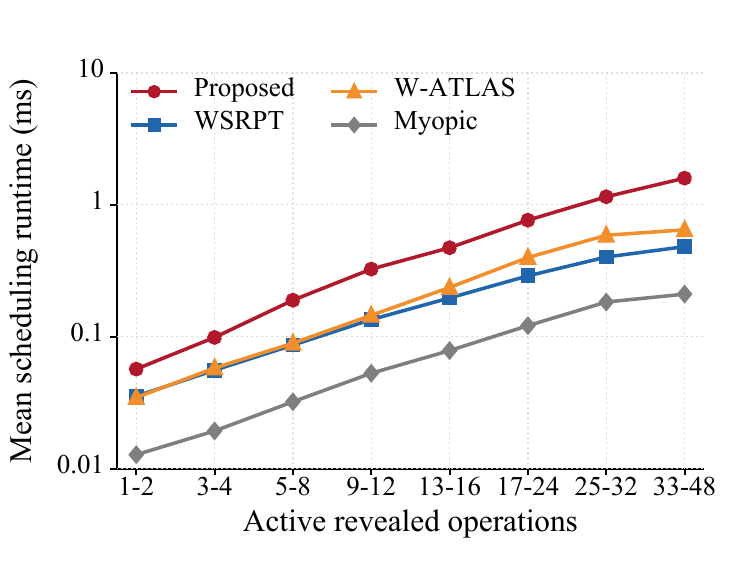}
\caption{Mean scheduling runtime versus active revealed operations.}
\label{fig:runtime_scalability}
\end{minipage}

\begin{minipage}[t]{0.318\textwidth}
\vspace{0pt}
\centering
\includegraphics[width=\linewidth]{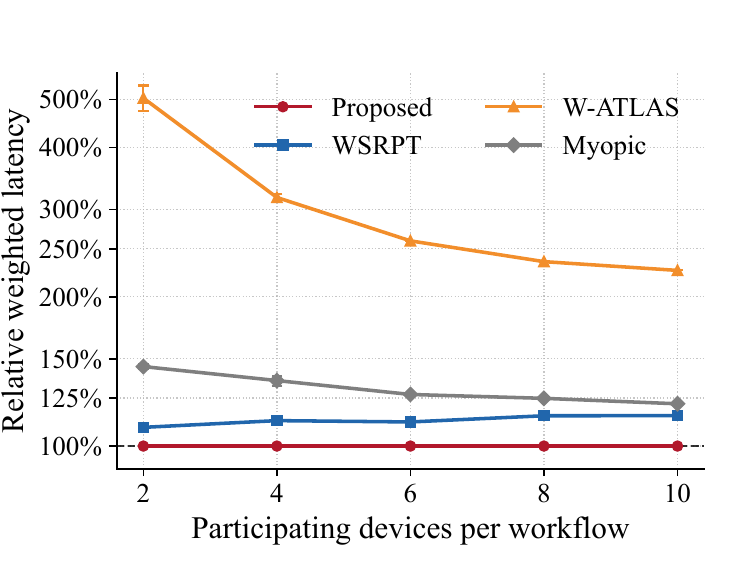}
\caption{Performance under different numbers of participating devices.}
\label{fig:device_scale}
\end{minipage}
\hfill
\begin{minipage}[t]{0.318\textwidth}
\vspace{0pt}
\centering
\includegraphics[width=\linewidth]{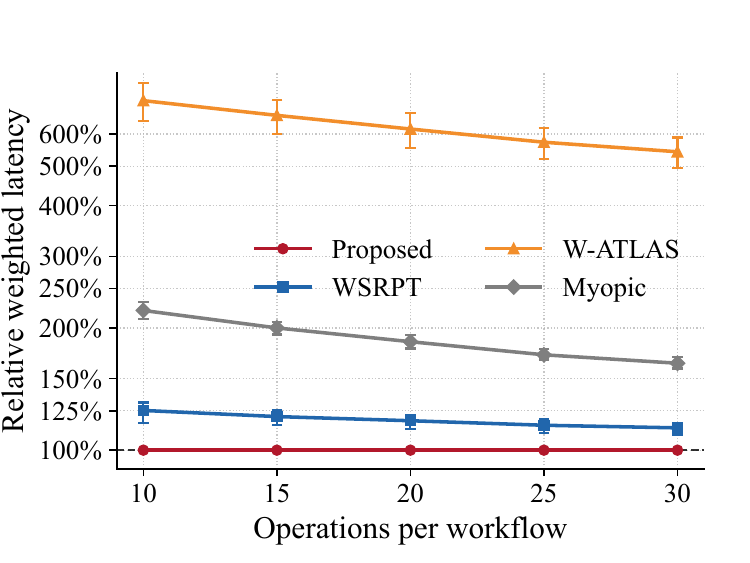}
\caption{Performance under different numbers of operations.}
\label{fig:workflow_size}
\end{minipage}
\hfill
\begin{minipage}[t]{0.318\textwidth}
\vspace{0pt}
\centering
\includegraphics[width=\linewidth]{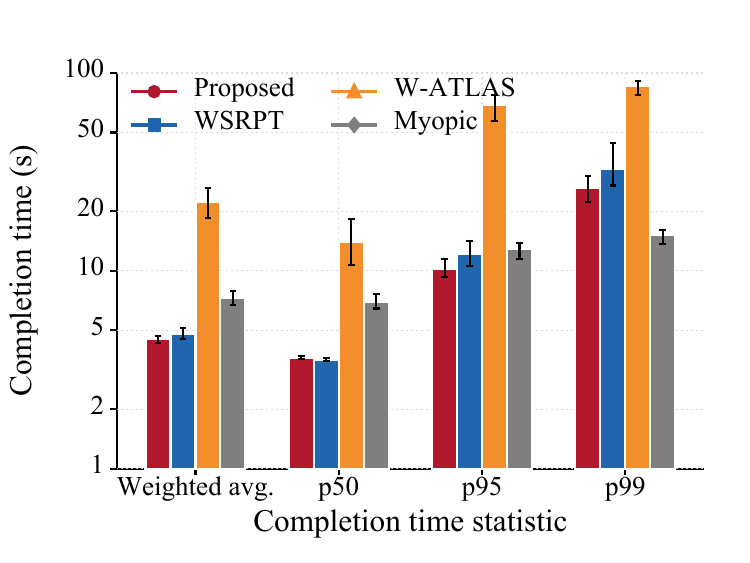}
\caption{Weighted completion time statistics.}
\label{fig:completion_time_summary}
\end{minipage}
\end{figure*}

\subsection{Simulation Setup}
In this section, we evaluate the proposed scheduling algorithm against three benchmarks under various settings.
Each job is represented by a hidden DAG,
while the scheduler only observes the operations revealed through corresponding reveal events.
A default workflow contains 12--40 operations,
with a depth of 4--12 and a maximum fan-out of four,
and selects communication endpoints from a pool of 20 devices.
Jobs arrive according to a Poisson process,
and their weights are sampled from $\{1,2,4\}$ with probabilities $\{0.6,0.3,0.1\}$.
Communication payloads and computation workloads follow bounded log-normal distributions.

Each slot lasts 10~ms.
The number of RBs available for agent traffic in each direction is independently and uniformly sampled from $[5,20]$.
The uplink and downlink transmission rates are 100 and 200~kbit per RB-slot, respectively.
For each device and transmission direction,
the achievable rate is scaled by an independent log-normal channel factor with a coefficient of variation of 0.1.
The edge server provides 3--6~GFLOP per slot, while the cloud execution is assumed to be contention-free,
providing 12~GFLOP per slot, with a one-way delay of 20~ms and 100/200-Mbps uplink/downlink backhaul.

The tail-latency predictor is trained using 1,000 independently generated workflows.
It estimates the remaining latency using the median latency indexed by task family,
operation type, and operation identifier, with the global training-set mean used as a fallback.
The reference communication and computation rates are updated using exponential moving averages with a coefficient of 0.2,
and the look-ahead parameter is set to $\zeta=25$.
The default arrival rate is set to 4 jobs/s, and each experiment contains 300 jobs.
To mitigate randomness, each data point is averaged over 10 independent random seeds.

We compare the proposed scheduler with WSRPT~\cite{jamalabadi2023wsrpt},
weighted ATLAS (W-ATLAS) adapted from the attained-service policy in Autellix~\cite{luo2025autellix},
and myopic equal-share scheduling.
For each revealed operation, W-ATLAS assigns a priority proportional to the
job weight divided by one plus the completed attained service along its revealed predecessor path.
It adopts the same myopic placement rule as WSRPT.

\subsection{Numerical Results}

Fig.~\ref{fig:load_performance} demonstrates the end-to-end scheduling performance under different traffic intensities.
As the job arrival rate increases from 1 to 6~jobs/s,
the weighted mean completion latency of Proposed increases from 3.08 to 9.23~s,
whereas that of WSRPT increases from 3.08 to 12.44~s.
At 6~jobs/s, Proposed reduces the latency by 25.58\%, 80.79\%, and 52.86\%
relative to WSRPT, W-ATLAS, and Myopic, respectively.
The widening performance gap under heavier traffic is consistent with the behavior of the baselines, 
which distribute service across active workflows without explicitly considering future impacts. 
Specifically, W-ATLAS may consistently favor newly arrived operations, 
while Myopic equal sharing ignores job weights and downstream workflow dependencies.

Fig.~\ref{fig:resource_regime} examines whether the benefit of joint scheduling persists as the system shifts between communication- and computation-limited regimes. 
WSRPT is the strongest non-Proposed baseline in every cell of the $5\times5$ grid. 
With five available RBs, the improvement of Proposed increases from 4.13\% at $0.5\times$ edge capacity to 
6.27\% at $2\times$ edge capacity, 
whereas with twenty RBs it is only 0.25\%--1.56\%. 
These results show that the gain is broadly consistent across the tested regimes but 
is largest when communication is the dominant bottleneck.

Fig.~\ref{fig:runtime_scalability} quantifies the online scheduling overhead as the number of active revealed operations increases. 
The plot contains eight bins from 1--2 through 33--48 operations, with at least 1,000 scheduler calls per algorithm in every retained bin. 
For Proposed, the mean runtime increases from 0.057~ms in the smallest bin to 1.594~ms in the largest.
At 33--48 operations, the corresponding means are 0.482~ms for WSRPT, 0.647~ms for W-ATLAS, and 0.211~ms for Myopic. 
The higher overhead of Proposed is consistent with its repeated evaluation of branch latencies and surrogate costs after each virtual completion. 
Nevertheless, its maximum runtime remains acceptable, and the runtime of all algorithms can be further reduced through implementations in more efficient programming languages and on more powerful hardware platforms.

Fig.~\ref{fig:device_scale} and Fig.~\ref{fig:workflow_size} 
illustrate the scheduling performance under different numbers of participating devices and workflow sizes, respectively. 
As the number of devices increases, the weighted mean completion latency of WSRPT remains between 109.08\% and 115.15\% 
relative to the proposed scheduler across the evaluated range. 
The latency ratio of W-ATLAS decreases from 503.38\% with two devices to 226.10\% with ten devices, 
while that of Myopic decreases from 144.66\% to 121.69\%. 
Thus, the proposed scheduler consistently achieves the lowest latency across all device scales. 
As the workflow size increases, the latency ratios of all baselines decrease: WSRPT decreases from 125.25\% to 113.48\%,
W-ATLAS from 725.37\% to 542.81\%, and Myopic from 220.76\% to 163.69\%, relative to the proposed scheduler.
These results show that Proposed maintains the lowest latency as either the number of participating devices or the workflow size increases,
with WSRPT remaining the closest benchmark.

Fig.~\ref{fig:completion_time_summary} presents the statistics of weighted completion latency. 
The weighted mean latency of Proposed is 4.50~s, compared with 4.81~s for WSRPT, 22.24~s for W-ATLAS, 
and 7.30~s for Myopic. 
Although WSRPT achieves a slightly lower weighted median latency than Proposed (3.54 versus 3.65~s), 
Proposed reduces the weighted p95 latency from 12.22 to 10.28~s and the weighted p99 latency from 32.61 to 26.23~s. 
This indicates that the advantage of Proposed over WSRPT mainly appears in the upper tail 
of the latency distribution rather than around the median. 
Myopic exhibits higher weighted mean, median, and p95 latencies than Proposed, 
but achieves a lower weighted p99 latency of 15.15~s. 
Therefore, a lower weighted mean latency does not necessarily imply consistent superiority across all tail quantiles.

\section{Conclusion}

This paper studied the joint scheduling of communication and computation for mobile edge AI agents whose workflows are revealed online.
By reformulating weighted completion time as a per-slot holding cost, we developed a workflow-aware scheduler that uses a look-ahead surrogate with completion correction to jointly determine computation placement and wireless and edge resource allocation.
Simulation results show that the proposed algorithm achieves the most significant reduction 
in weighted mean completion latency under heavy traffic and strong radio contention, while incurring limited runtime overhead. 

\bibliographystyle{IEEEtran}
\bibliography{ref}

\end{document}